\documentclass[12pt]{article}
\usepackage{graphicx}
\def\bra#1{\left\langle #1\right|}
\def\ket#1{\left| #1\right\rangle}
\newcommand{\bers}{\begin{eqnarray*}}
\newcommand{\eers}{\end{eqnarray*}}
\newcommand{\bt}{\begin{itemize}}
\newcommand{\et}{\end{itemize}}
\def\beq{\begin{equation}}
\def\eeq{\end{equation}}
\def\bea{\begin{eqnarray}}
\def\eea{\end{eqnarray}}
\def\nn{\nonumber}
\def\sss{\scriptscriptstyle}
\def\bd{B_d^0}
\def\bdbar{{\overline{B_d^0}}}
\def\bs{B_s^0}

\def\barp{{\raise.35ex\hbox
{${\sss (}$}}---{\raise.35ex\hbox{${\sss )}$}}}
\def\bdbarp{\hbox{$B_d$\kern-1.4em\raise1.4ex\hbox{\barp}}}
\def\bsbarp{\hbox{$B_s$\kern-1.4em\raise1.4ex\hbox{\barp}}}
\def\ks{K_{\sss S}}
\def\kbar{{\overline{K^0}}}
\def\roughly#1{\mathrel{\raise.3ex\hbox
{$#1$\kern-.75em\lower1ex\hbox{$\sim$}}}}
\def\lsim{\roughly<}

\def\mK{m_{\sss K}}

\def\ZI{Z_{\sss I}}
\def\ZR{Z_{\sss R}}
\def\tildeZI{{\tilde Z}_{\sss I}}
\def\tildeZR{{\tilde Z}_{\sss R}}
\def\calP{{\cal P}}
\newread\epsffilein 
\newif\ifepsffileok 
\newif\ifepsfbbfound 
\newif\ifepsfverbose 
\newdimen\epsfxsize 
\newdimen\epsfysize 
\newdimen\epsftsize 
\newdimen\epsfrsize 
\newdimen\epsftmp 
\newdimen\pspoints 
\def\epsfbox#1{\global\def\epsfllx{72}\global\def\epsflly{72}%
 \global\def\epsfurx{540}\global\def\epsfury{720}%
 \def\lbracket{[}\def\testit{#1}\ifx\testit\lbracket
 \let\next=\epsfgetlitbb\else\let\next=\epsfnormal\fi\next{#1}}%
\def\epsfgetlitbb#1#2 #3 #4 #5]#6{\epsfgrab #2 #3 #4 #5 .\\%
 \epsfsetgraph{#6}}%
\def\epsfnormal#1{\epsfgetbb{#1}\epsfsetgraph{#1}}%
\def\epsfgetbb#1{%
\openin\epsffilein=#1
\ifeof\epsffilein\errmessage{I couldn't open #1, will ignore it}\else
 {\epsffileoktrue \chardef\other=12
 \def\do##1{\catcode`##1=\other}\dospecials \catcode`\ =10
 \loop
 \read\epsffilein to \epsffileline
 \ifeof\epsffilein\epsffileokfalse\else
 \expandafter\epsfaux\epsffileline:. \\%
 \fi
 \ifepsffileok\repeat
 \ifepsfbbfound\else
 \ifepsfverbose\message{No bounding box comment in #1; using defaults}\fi\fi
 }\closein\epsffilein\fi}%
\def\epsfclipstring{}
\def\epsfsetgraph#1{%
 \epsfrsize=\epsfury\pspoints
 \advance\epsfrsize by-\epsflly\pspoints
 \epsftsize=\epsfurx\pspoints
 \advance\epsftsize by-\epsfllx\pspoints
 \epsfxsize\epsfsize\epsftsize\epsfrsize
 \ifnum\epsfxsize=0 \ifnum\epsfysize=0
 \epsfxsize=\epsftsize \epsfysize=\epsfrsize
 \epsfrsize=0pt
 \else\epsftmp=\epsftsize \divide\epsftmp\epsfrsize
 \epsfxsize=\epsfysize \multiply\epsfxsize\epsftmp
 \multiply\epsftmp\epsfrsize \advance\epsftsize-\epsftmp
 \epsftmp=\epsfysize
 \loop \advance\epsftsize\epsftsize \divide\epsftmp 2
 \ifnum\epsftmp>0
 \ifnum\epsftsize<\epsfrsize\else
 \advance\epsftsize-\epsfrsize \advance\epsfxsize\epsftmp \fi
 \repeat
 \epsfrsize=0pt
 \fi
 \else \ifnum\epsfysize=0
 \epsftmp=\epsfrsize \divide\epsftmp\epsftsize
 \epsfysize=\epsfxsize \multiply\epsfysize\epsftmp
 \multiply\epsftmp\epsftsize \advance\epsfrsize-\epsftmp
 \epsftmp=\epsfxsize
 \loop \advance\epsfrsize\epsfrsize \divide\epsftmp 2
 \ifnum\epsftmp>0
 \ifnum\epsfrsize<\epsftsize\else
 \advance\epsfrsize-\epsftsize \advance\epsfysize\epsftmp \fi
 \repeat
 \epsfrsize=0pt
 \else
 \epsfrsize=\epsfysize
 \fi
 \fi
 \ifepsfverbose\message{#1: width=\the\epsfxsize, height=\the\epsfysize}\fi
 \epsftmp=10\epsfxsize \divide\epsftmp\pspoints
 \vbox to\epsfysize{\vfil\hbox to\epsfxsize{%
 \ifnum\epsfrsize=0\relax
 \includegraphics{#1}%
 \else
 \epsfrsize=10\epsfysize \divide\epsfrsize\pspoints
 \includegraphics{#1}%
 \fi
 \hfil}}%
\global\epsfxsize=0pt\global\epsfysize=0pt}%
 {\catcode`\%=12 \global\let\epsfpercent=
\long\def\epsfaux#1#2:#3\\{\ifx#1\epsfpercent
 \def\testit{#2}\ifx\testit\epsfbblit
 \epsfgrab #3 . . . \\%
 \epsffileokfalse
 \global\epsfbbfoundtrue
 \fi\else\ifx#1\par\else\epsffileokfalse\fi\fi}%
\def\epsfempty{}%
\def\epsfgrab #1 #2 #3 #4 #5\\{%
\global\def\epsfllx{#1}\ifx\epsfllx\epsfempty
 \epsfgrab #2 #3 #4 #5 .\\\else
 \global\def\epsflly{#2}%
 \global\def\epsfurx{#3}\global\def\epsfury{#4}\fi}%
\def\epsfsize#1#2{\epsfxsize}
\pagestyle{plain}

\begin{document}

\begin{flushright}  
UdeM-GPP-TH-01-86\\
\end{flushright}
\vskip0.5truecm

\begin{center} 

{\large \bf
\centerline{Difficulties in Obtaining Weak Phases from $b\to d$
    Penguin Decays}}
\vspace*{1.0cm}
{\large Alakabha Datta$^{a,}$\footnote{email: datta@lps.umontreal.ca},
  C.S. Kim$^{b,}$\footnote{email: cskim@mail.yonsei.ac.kr,
    cskim@pheno.physics.wisc.edu} and David
  London$^{a,}$\footnote{email: london@lps.umontreal.ca}} \vskip0.3cm
{\it ${}^a$ Laboratoire Ren\'e J.-A. L\'evesque, Universit\'e de
  Montr\'eal,} \\
{\it C.P. 6128, succ.\ centre-ville, Montr\'eal, QC, Canada H3C 3J7} \\
\vskip0.3cm
{\it ${}^b$ Department of Physics and IPAP, Yonsei University, Seoul 120-749, Korea} \\
\vskip0.5cm
\bigskip
(\today)
\vskip0.5cm
{\Large Abstract\\}
\vskip3truemm
\parbox[t]{\textwidth} {We re-examine a recent proposal for obtaining
  $\beta$ using the measurements of the $b\to d$ penguin decays
  $\bd(t) \to K^0\kbar$ and $\bs(t) \to \phi \ks$, along with a
  theoretical assumption. We show that there are in fact three
  assumptions one can make, so that the method can in principle be
  used to extract $\alpha$, $\beta$ or $\gamma$. We also show that it
  is the assumption which yields $\gamma$ which is the best. However,
  the theoretical error on this assumption is still 25--30\%, which
  leads to an error on $\gamma$ of at least 20--25\%. Given our
  current understanding of hadronic physics, it does not seem possible
  to reduce this error.}
\end{center}
\thispagestyle{empty}
\newpage
\setcounter{page}{1}
\textheight 23.0 true cm
\baselineskip=14pt

Measurements of CP-violating rate asymmetries in the neutral $B$
system will allow one to obtain the CP angles $\alpha$, $\beta$ and
$\gamma$ \cite{BCPreview}. From these the unitarity triangle
\cite{PDG} can be constructed, and one can test the predictions of the
standard model (SM). If we are lucky, these measurements will reveal
the presence of physics beyond the SM.

One of the best ways of searching for new physics is to consider two
different decay modes whose CP asymmetries probe the same CP phase
within the SM. Any discrepancy between the measured values of these
asymmetries will point unequivocally to the presence of new physics.
One possibility is to consider pure $b\to d$ penguin decays such as
$\bd(t) \to K^0\kbar$ or $\bs(t) \to \phi \ks$. At the quark level,
these decay amplitudes take the form $b \to d {\bar s} s$. If such
decays are dominated by internal $t$-quark exchange, the amplitude is
proportional to $V_{tb} V_{td}^*$, where the $V_{ij}$ are elements of
the Cabibbo-Kobayashi-Maskawa (CKM) quark mixing matrix. In the
(approximate) Wolfenstein parametrization of the CKM matrix
\cite{Wolfenstein}, the only two matrix elements which have a nonzero
weak phase are $V_{td}$ [$\sim \exp(-i\beta)$] and $V_{ub}$ [$\sim
\exp(-i\gamma)$]. Thus, CP asymmetries in pure $b\to d$ penguin decays
probe the weak phase $\beta$. By comparing this value of $\beta$ with
that extracted via the conventional mode $\bd(t) \to J/\psi \ks$, one
can search for new physics.

Unfortunately, the $b\to d$ penguin amplitude is {\it not} dominated
by $t$-quark exchange. For example, the $u{\bar u}$ quark pair of the
tree-level decay $b\to u {\bar u} d$ can rescatter strongly into an $s
{\bar s}$ quark pair, giving an effective $V_{ub} V_{ud}^*$
contribution to the $b\to d$ penguin amplitude, and similarly for the
$b\to c {\bar c} d$ tree-level decay. Buras and Fleischer
\cite{ucquark} have noted that the $u$- and $c$-quark contributions
can be between 20\% and 50\% of the leading $t$-quark contribution to
the $b\to d$ penguin amplitude. And since $V_{ub} V_{ud}^*$ and
$V_{cb} V_{cd}^*$ have different weak phases as compared to $V_{tb}
V_{td}^*$, this implies that CP asymmetries in pure $b\to d$ penguin
decays do not cleanly probe the weak phase $\beta$.

But this then begs the question: is it possible to isolate the
$t$-quark contribution to the $b\to d$ penguin amplitude? If so, one
could then obtain $\beta$ from this piece of the amplitude, and
compare it with the value found in $\bd(t) \to J/\psi \ks$.
Unfortunately, as was shown in Ref.~\cite{LSS}, this is not possible.
In a nutshell, the argument is as follows. The $b\to d$ penguin
amplitude can be written generally as
\beq
P = P_u V_{ub} V_{ud}^* + P_c V_{cb} V_{cd}^* + P_t V_{tb} V_{td}^* ~.
\eeq
Now, the unitarity of the CKM matrix implies that $V_{ub} V_{ud}^* +
V_{cb} V_{cd}^* + V_{tb} V_{td}^* = 0$. Thus, any of the three pieces
in the above amplitude can always be eliminated in terms of the
remaining two. For example, if we eliminate the $V_{ub} V_{ud}^*$
piece, we obtain
\bea
P & = & (P_c - P_u) V_{cb} V_{cd}^* + (P_t - P_u) V_{tb} V_{td}^* \nn \\
& \equiv & \calP_{cu} \, e^{i\delta_{cu}} + \calP_{tu} \,
e^{i\delta_{tu}} \, e^{i\beta} ~,
\label{param1}
\eea
where we have explicitly separated out the weak and strong phases and
absorbed the magnitudes $|V_{cb}^* V_{cd}|$ and $|V_{tb}^* V_{td}|$
into the definitions of $\calP_{cu}$ and $\calP_{tu}$, respectively.
Similarly, eliminating $V_{tb} V_{td}^*$ gives
\beq
P = \calP_{ct} \, e^{i\delta_{ct}} + \calP_{ut} \, e^{i\delta_{ut}} \,
e^{-i\gamma} ~.
\label{param2}
\eeq
Now, suppose that a method existed which would permit one to extract
the CP phase $\beta$, with no hadronic uncertainties, using the
parametrization of Eq.~(\ref{param1}). If so, then $\beta$ could be
expressed entirely in terms of experimentally measured quantities.
However, Eq.~(\ref{param2}) has the same form as Eq.~(\ref{param1}).
Thus, this same method would allow us to cleanly obtain $-\gamma$
using Eq.~(\ref{param2}). In particular, $-\gamma$ would be expressed
as the same function of observables as $\beta$, leading to the
conclusion that $\beta = -\gamma$. Since this is clearly not true in
general, one concludes that it is not possible to cleanly measure the
weak phase of the $t$-quark piece (or indeed any other piece) of the
$b\to d$ penguin amplitude. In Ref.~\cite{LSS}, this is referred to as
the ``CKM ambiguity.''

However, in Ref.~\cite{LSS} it is also argued that it is possible to
resolve the CKM ambiguity, and hence isolate a particular piece of the
$b\to d$ penguin amplitude, if one makes an assumption regarding the
hadronic parameters involved in this amplitude. This fact has been
used by two of the present authors (Kim, London) and Yoshikawa in
Ref.~\cite{KLY} to obtain $\beta$ from the $t$-quark piece of the
$b\to d$ penguin amplitude. The idea is to use the two penguin decays
$\bd(t) \to K^0\kbar$ and $\bs(t) \to \phi \ks$. If one eliminates the
$V_{ub} V_{ud}^*$ piece from the penguin amplitudes, as in
Eq.~(\ref{param1}), the amplitudes for these two decays can be written
as
\bea
A(\bd\to K^0\kbar) & = & \calP_{cu} \, e^{i\delta_{cu}} + \calP_{tu}
\, e^{i\delta_{tu}} \, e^{-i\beta} ~, \nn\\
A(\bs\to\phi\ks) & = & \calP'_{cu} \, e^{i\delta'_{cu}} + \calP'_{tu}
\, e^{i\delta'_{tu}} \, e^{-i\beta} ~.
\label{BKKamps}
\eea
The following assumption is now made:
\beq
{r_u \over r_u'} \equiv { \calP_{cu}/\calP_{tu} \over
  \calP'_{cu}/\calP'_{tu} } = 1 ~.
\label{uassumption}
\eeq
(Note that the dependence on the CKM matrix elements cancels in this
ratio, so that this really is an assumption about the hadronic
parameters of the two amplitudes.) Measurements of $\bd(t) \to
K^0\kbar$ and $\bs(t) \to \phi \ks$, combined with the assumption in
Eq.~(\ref{uassumption}), will then allow the extraction of $\beta$,
the weak phase of the $t$-quark piece of the $b\to d$ penguin
amplitude. By comparing this value of $\beta$ with that found in
$\bd(t) \to J/\psi \ks$, one may be able to detect the presence of new
physics. In what follows, we will refer to this as the ``KLY method.''

However, there is a problem with this method which has been overlooked
in Ref.~\cite{KLY}. It is similar to the CKM ambiguity: how do we know
that $V_{ub} V_{ud}^*$ is the correct term to eliminate? For example,
had we eliminated the $V_{tb} V_{td}^*$ piece, as in
Eq.~(\ref{param2}), the amplitudes would take the form
\bea
A(\bd\to K^0\kbar) & = & \calP_{ct} \, e^{i\delta_{ct}} + \calP_{ut}
\, e^{i\delta_{ut}} \, e^{i\gamma} ~, \nn\\
A(\bs\to\phi\ks) & = & \calP'_{ct} \, e^{i\delta'_{ct}} + \calP'_{ut}
\, e^{i\delta'_{ut}} \, e^{i\gamma} ~.
\eea
Using the assumption that
\beq
{r_t \over r_t'} \equiv { \calP_{ct}/\calP_{ut} \over
  \calP'_{ct}/\calP'_{ut} } = 1 ~,
\label{tassumption}
\eeq
the KLY method can be used to allow us to obtain $\gamma$. Similarly,
the elimination of the $V_{cb} V_{cd}^*$ piece in the amplitudes,
combined with the assumption that
\beq
{r_c \over r_c'} \equiv { \calP_{uc}/\calP_{tc} \over
  \calP'_{uc}/\calP'_{tc} } = 1 ~,
\label{cassumption}
\eeq
would permit the extraction of $\alpha$. The confusion regarding the
validity of the three assumptions in Eqs.~(\ref{uassumption}),
(\ref{tassumption}) and (\ref{cassumption}) can be thought of as a
``second CKM ambiguity.''

It is clear that the three assumptions in Eqs.~(\ref{uassumption}),
(\ref{tassumption}) and (\ref{cassumption}) cannot be simultaneously
true. If they were, then, as was the case in the discussion of the CKM
ambiguity, one would obtain results such as $\beta = -\gamma$, which
do not hold in general. We therefore deduce that (at least) two of the
assumptions are poor. That is, the values of the CP phases obtained
using these assumptions will differ enormously from their true values.

In fact, it can be argued that all three assumptions are likely to be
poor. After all, even though they are based on the same quark-level
process, the hadronic quantities (form factors, strong phases, etc.)
describing the decays $\bd(t) \to K^0\kbar$ and $\bs(t) \to \phi \ks$
should be quite different. Thus, there is no reason to expect any of
the ratios in Eqs.~(\ref{uassumption}), (\ref{tassumption}) and
(\ref{cassumption}) to equal one, in which case it appears that the
method is of little practical use.

However, this is not as serious a problem as appears at first glance.
Consider again the assumption of Eq.~(\ref{uassumption}). In fact, the
KLY method does not require that $r_u/r'_u = 1$. All that is necessary
is that we know the value of this ratio. Thus, if we could
theoretically calculate the value of $r_u/r'_u$, the KLY method could
then be used to obtain $\beta$. Similarly, if we could calculate
$r_t/r'_t$ or $r_c/r'_c$, we could extract $\gamma$ or $\alpha$. We
therefore see that the second CKM ambiguity is not necessarily a
disadvantage: depending on what hadronic information is known, any of
the three CP phases could, in principle, be obtained using the KLY
method.

This is therefore the key question: how well can we estimate the
values of the three ratios $r_u/r'_u$, $r_t/r'_t$ and
$r_c/r'_c$? Can we make general statements regarding this question,
or is it process-dependent? Also, for a given theoretical uncertainty
on a particular ratio, what is error on the corresponding extracted CP
phase? These are the issues which we address in this paper.

We argue below that, with our present theoretical understanding of
hadronic $B$ decays, neither $r_u/r'_u$ nor $r_c/r'_c$ can be computed
with any degree of reliability. Therefore the KLY method cannot be
used to obtain $\beta$ or $\alpha$. On the other hand, for pure $b\to
d$ penguin decays such as $\bd(t) \to K^0\kbar$ and $\bs(t) \to \phi
\ks$, $r_t/r'_t$ can be calculated to lie in a much narrower
range. Thus, for these decays, the KLY method can be used in principle
to extract $\gamma$. Unfortunately, as we will see, the error on
$\gamma$ remains fairly large, in the range of 20--25\%, so that this
method cannot be used to obtain a precise measurement of $\gamma$. We
examine possible ways to reduce this error.

We begin our analysis by addressing the question of how to estimate
the ratios $r_u/r'_u$, $r_t/r'_t$ and $r_c/r'_c$. The important point
to realize is that the $P_i$'s which appear in the ratios
[Eqs.~(\ref{uassumption}), (\ref{tassumption}), (\ref{cassumption})]
are actually matrix elements of penguin operators. Thus, in order to
answer this question, we need some sort of framework in which to
evaluate hadronic matrix elements. Factorization is usually employed
to calculate nonleptonic amplitudes. Corrections to naive
factorization have been calculated within QCD factorization
\cite{BBNS} and the perturbative QCD (pQCD) \cite{PQCD} approach. In
QCD factorization, naive factorization is recovered as the
leading-order term, and one systematically computes corrections to it
in an expansion in $\alpha_s(m_b) \sim 0.2$ and $\Lambda_{QCD}/m_b$.
At $O(\alpha_s)$ the vertex and hard spectator corrections modify only
the top penguin amplitude and do not introduce any additional weak
phases. The penguin or the rescattering corrections generate the up
and the charm penguin pieces. The SM effective hamiltonian for
hadronic $B$ decays is \cite{Buras}:
\begin{eqnarray}
H_{eff}^q &=& {G_F \over \protect \sqrt{2}}
[V_{fb}V^*_{fq}(c_1O_{1f}^q + c_2 O_{2f}^q) \nn\\ 
& & \qquad\qquad - \sum_{i=3}^{10}(V_{ub}V^*_{uq} c_i^u
+V_{cb}V^*_{cq} c_i^c +V_{tb}V^*_{tq} c_i^t) O_i^q] + h.c.
\label{H_eff}
\end{eqnarray}
Here, $q$ can be either a $d$ or an $s$ quark, depending on whether
the decay is a $\Delta S = 0$ or a $\Delta S = -1$ process. In the
first terms, $f$ can be a $u$ or a $c$ quark, while in the last terms,
the superscript $u$, $c$ or $t$ indicates the flavour of the internal
quark. Note that, in the effective Hamiltonian $H_{eff}$ above, we
have explicitly included the up and charm penguin pieces which are
generated by rescattering.  The values of the Wilson coefficients
$c_i^f$ for the penguin operators evaluated at the scale $\mu = m_b =
5$ GeV, for $m_t = 176$ GeV and $\alpha_s(m_Z) = 0.117$, are
\cite{FSHe}:
\begin{eqnarray}
c^t_3 = 0.017 ~,~~ c^t_4 = -0.037 &,& c^t_5 = 0.010 ~,~~
c^t_6 =-0.045 ~, \nn\\
c^t_7 = -1.24\times 10^{-5} ~,~~ c_8^t = 3.77\times 10^{-4} &,&
c_9^t = -0.010 ~,~~ c_{10}^t = 2.06\times 10^{-3} ~, \nn\\
c_{3,5}^i = -c_{4,6}^i/N_c = P^i_s/N_c &,&
c_{7,9}^i = P^i_e ~,~~
c_{8,10}^i = 0 ~,~~
i=u,c ~,
\label{coeffs}
\end{eqnarray}
where $N_c$ is the number of colors. The leading contributions to
$P^i_{s,e}$ are given by $P^i_s = ({\frac{\alpha_s}{8\pi}}) c_2
({\frac{10}{9}} +G(m_i,\mu,q^2))$ and $P^i_e =
({\frac{\alpha_{em}}{9\pi}}) (N_c c_1+ c_2) ({\frac{10}{9}} +
G(m_i,\mu,q^2))$, where the function $G(m,\mu,q^2)$ takes the form
\begin{eqnarray}
G(m,\mu,q^2) = 4\int^1_0 x(1-x) \mbox{ln}{m^2-x(1-x)q^2\over
\mu^2} ~\mbox{d}x ~,
\end{eqnarray}
where $q$ is the momentum carried by the virtual gluon in the penguin
diagram.

Of course, as mentioned above, we are really interested in the matrix
elements of the various operators for the decay $B \to f_1 f_2$. We
therefore define new coefficients ${\bar{c}}_i^{u,c}$ as
\beq
{\bar{c}}_i^{u,c} 
= \frac{\bra{f_1 f_2}c_i^{u,c}(q^2)O_i\ket{B}}
{\bra{f_1 f_2}O_i\ket{B}} ~.
\eeq
The values of ${\bar{c}}_i^{u,c}$ can be calculated in the approaches
of Ref.~\cite{BBNS} and Ref.~\cite{PQCD} if the light cone
distributions of the various mesons are known.  However, the values
for the ${\bar{c}}_i^{u,c}$ will, in general, be different in the QCD
factorization and the pQCD calculations.  This is because pQCD assumes
that the the light quarks forming the final state light mesons must
all be energetic while QCD factorization assumes that the spectator
quark coming from the $B$ mesons remains soft as it combines with an
energetic quark to form one of the final-state light mesons.  Without
adopting a particular approach, we will follow the usual practice,
which is to simply replace
\beq
{\bar{c}}_i^{u,c} \to c_i^{u,c}(q^2_{av}) ~,
\label{cbarc}
\eeq
and we will conservatively allow $q^2_{av}$ to vary between $m_b^2/4
\to m_b^2/2$. (Note that this is the range which has been used in the
past to take into account possible process dependence \cite{Desh,
WylerPalmer}.) With this prescription, the relations between the
various ${\bar{c}}_i^{u,c}$ are the same as those between the various
$c_i^{u,c}(q^2)$. In particular, we still have \cite{dkl2}
\beq
{\bar c}_4^{u,c} = {\bar c}_6^{u,c} ~.
\label{cbarrel}
\eeq

In order to calculate matrix elements, we need to consider specific
final states. Consider first the decay $\bd \to K^0 \bar{K}^0$. Using
the naive factorization approximation, along with the fact that
$\bar{c}_6^{u,c}= \bar{c}_4^{u,c}$ [Eq.~(\ref{cbarrel})], one can
write
\beq
P_{u,c} = \bar{c}_6^{u,c} (1-\frac{1}{N_c^2})
\left[\left\langle{O_{LL}}\right\rangle
-2\left\langle{O_{SP}}\right\rangle \right] ~,
\label{Puceqn}
\eeq
where
\bea
\left\langle{O_{LL}}\right\rangle & = &
\bra{\bar{K}^0}\bar{s}\gamma_{\mu}(1-\gamma_5)b\ket{\bd}
\bra{K^0}\bar{d}\gamma^{\mu}(1-\gamma_5)s\ket{0} ~, \nn\\
\left\langle{O_{SP}}\right\rangle & = &
\bra{\bar{K}^0}\bar{s}(1-\gamma_5)b\ket{\bs}
\bra{K^0}\bar{d}(1+\gamma_5)s\ket{0} ~,
\label{ucquark}
\eea
and we have dropped factors common to $P_{u,c,t}$. (The operator
$O_{SP}$ appears due to a Fierz transformation: $(V-A) \otimes (V+A) =
-2 (S-P) \otimes (S+P)$.) The contribution from the top penguin
amplitude is given by
\bea
P_t & = &\left[(c_4^t +\frac{c_3^t}{N_c})-\frac{1}{2}
(c_{10}^t +\frac{c_9^t}{N_c})\right]
\left\langle{O_{LL}}\right\rangle 
-2(c_6^t +\frac{c_5^t}{N_c})\left\langle{O_{SP}}\right\rangle ~.
\eea
In the above, we have neglected the contributions from $c_{7,8}$. 

It is convenient to rewrite $P_{u,c}$ and $P_t$ as
\bea
P_{u,c} & = & \bar{c}_6^{u,c} (1-\frac{1}{N_c^2})X_r
\left\langle{O_{LL}}\right\rangle ~, \nn\\
P_t & = & 
+ (a_4-a_6-\frac{1}{2}a_{10})\left\langle{O_{LL}}\right\rangle +
a_6 X_r\left\langle{O_{LL}}\right\rangle  ~,
\label{samp}
\eea
where
\beq
a_i = \cases{ c_i + {c_{i-1} \over N_c} ~, & $i = 4,6,10$ ~,\cr
              c_i + {c_{i+1} \over N_c} ~, & $i = 3,5,9$ ~,\cr}
\label{tquark2}
\eeq
and
\beq
X_r   \equiv   \left[ 1 - {2 \left\langle{O_{SP}}\right\rangle \over
    \left\langle{O_{LL}}\right\rangle} \right]  ~.
\label{Xdef}
\eeq
{}For $\bd \to
K^0 \bar{K^0}$,
\beq
X_r = \left[ 1 + {2 \mK^2 \over m_b } \, {1 \over m_s + m_d}
\right] ~.
\label{XKK}
\eeq
Note that, for $\mK = 500$ MeV, $m_b \simeq 5$ GeV, $m_s \simeq 100$
MeV and $m_d \simeq 0$, one finds that $X_r \simeq 2$ for this decay. 

Another possible decay mode is $\bd \to K^* \bar{K}^*$. In this case
the above analysis is unchanged, except that
\beq
X_r=1 ~,
\eeq
since here $\langle{O_{SP}}\rangle = 0$. (Note that the KLY method
does not apply to the decay $\bd \to {\bar K} K^*$ since this final
state is not CP self-conjugate.) Finally, one can also consider decays
such as $\bs\to\phi\ks$. However, over a large region of parameter
space, $P_u = P_c \simeq 0$ \cite{dkl2}, so that the CP asymmetry
probes $\beta$ directly, and the KLY method does not apply. We
therefore concentrate only on $\bd \to K^{(*)} \bar{K}^{(*)}$ decays
in the analysis below.

One can now construct the ratios $r_u$, $r_c$ and $r_t$ defined in the
numerators of Eqs.~(\ref{uassumption}), (\ref{tassumption}) and
(\ref{cassumption}). We use $r_{u0}$, $r_{c0}$ and $r_{t0}$ to denote
the values of $r_u$, $r_c$ and $r_t$ in the naive factorization limit.
In addition, it is useful to separate the dependence of $r_{u0}$,
$r_{c0}$ and $r_{t0}$ on the CKM matrix elements from the hadronic
dependence:
\beq
r_{u0} = \left| V_{cb}^* V_{cd} \over V_{tb}^* V_{td} \right|
r_{u0}^{(had)} ~,~~ 
r_{c0} = \left| V_{ub}^* V_{ud} \over V_{tb}^* V_{td} \right|
r_{c0}^{(had)} ~,~~
r_{t0} = \left| V_{cb}^* V_{cd} \over V_{ub}^* V_{ud} \right|
r_{t0}^{(had)} ~.
\eeq
Using the above expressions for the $P_i$, we find
\bea
r_{u0}^{(had)} & = & \left|\frac{P_c-P_u}{P_t-P_u}\right| =
\left|\frac{X_r(1-N_c^2) \left(\bar{c}_6^{c} -\bar{c}_6^{u}\right) }
  {a_4+a_6(X_r-1)-\frac{1}{2}a_{10}
    -X_r(1-N_c^2)\bar{c}_6^{u} }\right| ~, \nn\\
r_{c0}^{(had)} & = & \left|\frac{P_u-P_c}{P_t-P_c}\right| =
\left|\frac{X_r(1-N_c^2) \left(\bar{c}_6^{u}
      -\bar{c}_6^{c}\right) }
  {(a_4+a_6(X_r-1)-\frac{1}{2}a_{10})-X_r(1-N_c^2)\bar{c}_6^c}
\right| ~, \nn\\
r_{t0}^{(had)} & = & \left|\frac{P_c-P_t}{P_u-P_t}\right| =
\left|\frac{(a_4+a_6(X_r-1)-\frac{1}{2}a_{10})-
    X_r(1-N_c^2)\bar{c}_6^c}
  {(a_4+a_6(X_r-1)-\frac{1}{2}a_{10})-X_r(1-N_c^2)\bar{c}_6^u}
\right| ~.
\label{r0}
\eea
One obtains similar expressions for the ratios $r_{u0}^{\prime(had)}$,
$r_{c0}^{\prime(had)}$ and $r_{t0}^{\prime(had)}$, where the $r'_i$
($i=u,c,t$) are defined in the denominators of
Eqs.~(\ref{uassumption}), (\ref{tassumption}) and (\ref{cassumption}).
Note that the various ratios $r_{u0,c0,t0}^{(had)}$ and
$r_{u0,c0,t0}^{\prime(had)}$ depend only on the Wilson coefficients,
on $q^2_{av}$ and on the quark masses in the factor $X_r$. There is no
dependence on hadronic quantities such as form factors and decay
constants since they cancel in the ratios.

So far we have considered only the rescattering corrections which
occur at $\alpha_s$, and which generate the up and charm penguins.
However, as noted earlier, there are additional vertex and the hard
spectator corrections, which are also $O(\alpha_s)$. These can be
taken into account by the replacement $a_i \to a_i^{eff} =a_i(1+t_i)$
in Eq.~(\ref{tquark2}), where $t_i \sim O(\alpha_s)$ are
process-dependent corrections to the naive factorization assumption.
Since the corrections $t_i$ depend on several poorly-known
nonperturbative quantities, we will treat them as free parameters.
The process dependence of the $r_{u,c,t}^{(had)}$ and
$r_{u,c,t}^{\prime(had)}$ then comes from three sources: (i) the value
of the momentum transfer $q^2_{av}$ in Eq.~(\ref{cbarc}), (ii) $t_i$,
the $O(\alpha_s)$ corrections to the $a_i$, and (iii) the dependence
of the quantity $X_r$ [Eq.~(\ref{Xdef})] on the quark and hadron
masses.

Including now all the corrections to naive factorization, to first
order in $\alpha_s$ and to leading order in $\Lambda_{QCD}/m_b$, we
can make the replacement $P_t \to P_t(1+x)$, where the
process-dependent quantity $x \sim \alpha_s(m_b) \sim 0.2$
\footnote{We have neglected a possible small complex phase in
$x$.}. We can then obtain the corrected values of $r_u^{(had)}$,
$r_c^{(had)}$ and $r_t^{(had)}$ as
\bea
r_u^{(had)} & = & r_{u0}^{(had)} \, {1 \over \sqrt {1 + x^2 {|P_t|^2
      \over |P_{tu}|^2}
    + 2 x {|P_t| \over |P_{tu}|} \cos(\delta_{tu}-\delta_t)}} ~, \nn\\
r_c^{(had)} & = & r_{c0}^{(had)} \, {1 \over \sqrt {1 + x^2 {|P_t|^2
      \over |P_{tc}|^2}
    + 2 x {|P_t| \over |P_{tc}|} \cos(\delta_{tc}-\delta_t)}} ~, \nn\\
r_t^{(had)} & = & r_{t0}^{(had)} \, \sqrt{ { 1 + x^2 {|P_t|^2 \over
      |P_{tc}|^2} + 2 x {|P_t| \over |P_{tc}|}
    \cos(\delta_{tc}-\delta_t) \over 1 + x^2 {|P_t|^2 \over
      |P_{tu}|^2} + 2 x {|P_t| \over |P_{tu}|}
    \cos(\delta_{tu}-\delta_t)}} ~,
\label{rcorrect}
\eea
where $P_t - P_u = |P_t| e^{i\delta_t} - |P_u| e^{i\delta_u} \equiv
|P_{tu}| e^{i\delta_{tu}}$, and similarly for $P_t - P_c$. As usual,
there are similar expressions for $r_u^{\prime(had)}$,
$r_c^{\prime(had)}$ and $r_t^{\prime(had)}$. From these expressions
one can see that the nonfactorizable corrections tend to cancel in
$r_t^{(had)}$ and $r_t^{\prime(had)}$, and so these ratios are the
least affected by such effects. Note also that, as mentioned
previously, the dependence on the CKM matrix elements cancels in the
ratios $r_i/r'_i$, i.e.\ 
\beq
{r_i\over r'_i} = {r_i^{(had)} \over r_i^{\prime(had)}} ~.
\eeq

We are now in a position to calculate the theoretically-allowed ranges
of the three ratios of Eqs.~(\ref{uassumption}), (\ref{tassumption})
and (\ref{cassumption}). There are several factors which can
contribute to these ranges: since the quantities $r_i^{(had)}$ and
$r_i^{\prime(had)}$ ($i=u$, $c$, $t$) are calculated for different
processes, the momentum transfer $q^2_{av}$, the parameter $X_r$
[Eq.~(\ref{Xdef})], and the nonfactorizable correction $x$ may all be
different. We therefore adopt the following procedure. For each of
$r_i^{(had)}$ and $r_i^{\prime(had)}$, we allow the values of the
quark masses to vary in the following ranges: $4.3 \le m_b \le 4.9$
GeV, $1.20 \le m_c \le 1.30$ GeV, and $0.080 \le m_s \le 0.120$ GeV.
We fix $m_d=6$ MeV. All masses are taken to be at the $b$-quark mass
scale. In addition, for a given value of $m_b$, we vary $q^2_{av}$
between $m_b^2/4$ and $m_b^2/2$. Finally, we take $x$ to be in the
range $-0.2 \le x \le 0.2$.

In order to calculate the $r_i^{(had)}$ and $r_i^{\prime(had)}$, we
must choose two decay processes. As a first example, we consider
$\bd(t) \to K^*(892) {\bar K}^*(892)$ and $\bd(t) \to
K^*(1410) {\bar K}^*(1410)$. The quantum numbers ($J^{PC}$) of
the $K^*(892)$ and $K^*(1410)$ are the same, and the latter can
be interpreted as a radially excited $K^*$. From the analysis of
Ref.~\cite{DattaLipkin} we can expect the branching ratio of $\bd \to
K^*(1410) {\bar K}^*(1410)$ to be comparable or enhanced relative
to $\bd \to K^*(892) {\bar K}^*(892)$ \footnote{It may also be
  useful to consider the process $\bd(t) \to K^*(1680)
  {\bar K}^*(1680)$, since the $K^*(1680)$ has a significantly
  higher branching ratio to $K \pi$ ($\sim 39$\%) than the $K^*(1410)$
  ($\sim 7$ \%). This makes reconstructing the $K^*(1680)
  \bar{K^*}(1680)$ final state easier than $K^*(1410)
  {\bar K}^*(1410)$.}. (Note that since the final state consists of
two vector mesons, an angular analysis will have to be performed to
separate the three helicity states \cite{helicity}. However, that does
not affect our analysis here.) We find that $r_u^{(had)}$ lies in the
range $0.14 \le r_u^{(had)} \le 0.54$, and similarly for
$r_u^{\prime(had)}$. Thus, we have $0.26 \le r_u/r_u' \le 3.86$.
Similarly, the range of $r_c^{(had)}$ and $r_c^{\prime(had)}$ is from
0.13 to 0.47, which leads to $0.28 \le r_c/r_c' \le 3.62$. The allowed
ranges for both $r_u/r_u'$ and $r_c/r_c'$ are clearly enormous. Since
the KLY method requires a reasonably accurate theoretical prediction
of these ratios, we therefore conclude that $\beta$ and $\alpha$
cannot be obtained using this method.

On the other hand, the allowed range for $r_t^{(had)}$ and
$r_t^{\prime(had)}$ is considerably narrower: $0.92 \le
r_t^{(had)},~r_t^{\prime(had)} \le 1.23$. This is due partly to the
fact that $P_u$ and $P_c$ are both quite a bit smaller than $P_t$, so
that the numerator and denominator of $r_t^{(had)}$ are roughly equal
(and similarly for $r_t^{\prime(had)}$), and partly to the fact that
the nonfactorizable corrections approximately cancel in $r_t^{(had)}$
and $r_t^{\prime(had)}$ [Eq.~(\ref{rcorrect})]. We therefore find
that $0.75 \le r_t/r_t' \le 1.34$, a considerably tighter range than
that found for $r_u/r_u'$ and $r_c/r_c'$. Thus, of the three ratios,
$r_t/r_t'$ has by far the narrowest range, so that in fact it is the
CP phase $\gamma$ which can be extracted with the smallest error using
the KLY method. Note also that the average value of $r_t/r_t'$ in its
range is 1.04, which is quite close to unity. Thus, the assumption of
Eq.~(\ref{tassumption}) is justified, though of course the key
question is the error on the assumption. {}From now on, we therefore
consider only the measurement of $\gamma$ using the assumption of
Eq.~(\ref{tassumption}).

Another pair of processes that one can consider are $\bd(t) \to K^{0}
\bar{K}^{0}$ and $\bd(t) \to K^{0}(1460) \bar{K}^{0}(1460)$
($K^0(1460)$ has the same quantum numbers as the $K^0$). Since these
final states consist of two pseudoscalars, one must use the expression
for $X_r$ found in Eq.~(\ref{XKK}). However, the results do not change
much: we find $0.94 \le r_t^{(had)},~r_t^{\prime(had)} \le 1.22$,
leading to $0.77 \le r_t/r_t' \le 1.30$.

Note that we have been extremely generous in estimating the allowed
range or $r_t/r_t'$. We have allowed each of $r_t^{(had)}$ and
$r_t^{\prime(had)}$ to take any value in their allowed ranges. That
is, we have assumed that the momentum transfer $q^2_{av}$ and the
nonfactorizable correction $x$ can each take completely different
values in the two decay processes. However, in practice, this is not
likely to be the case if two similar final states are chosen. For
example, we expect a similar value of $q^2_{av}$ in the two decays
$\bd(t) \to K^*(1410) {\bar K}^*(1410)$ and $\bd(t) \to
K^{0}(1460) \bar{K}^{0}(1460)$, since the masses of the particles in
the final states are almost equal. Also, it is reasonable to expect
that the effect of nonfactorizable contributions will be similar for
the decays $\bd(t) \to K^*(892) {\bar K}^*(892)$ and $\bd(t) \to
K^*(1410) {\bar K}^*(1410)$, since both final states consist of
two vector mesons. Thus, it is likely that the range of $r_t/r_t'$ may
actually be much smaller than that calculated above, particularly for
similar final states. We will come back to this point later.

Of course, it is not enough to have established that it is $\gamma$
which can be extracted with the smallest error using the KLY method.
What we really want to know is: what is the size of the theoretical
error on $\gamma$ in this method? This is the question we now address.

We first recall how measurements of two processes, $\bd(t) \to M_1
M_2$ and $\bd(t) \to M'_1 M'_2$, along with an assumption about the
ratio of penguin amplitudes, can be used to obtain $\gamma$. Using the
convention $\bd = {\bar b}d$, we write the amplitude for $\bd \to
M_1M_2$ using the parametrization of Eq.~(\ref{param2}):
\beq
A^{M_1M_2}_d = \calP_{ct} \, e^{i\delta_{ct}} + \calP_{ut} \,
e^{i \delta_{ut}} \, e^{i\gamma} ~.
\label{amp}
\eeq
The amplitude for $\bd \to M'_1 M'_2$ can be written similarly:
\beq
A^{M'_1 M'_2}_d = \calP'_{ct} \, e^{i \delta'_{ct}} + {\cal
  P}'_{ut} \, e^{i \delta'_{ut}} \, e^{i\gamma} ~.
\label{amptilde}
\eeq
The amplitudes for $\bdbar\to M_1 M_2$ and $\bdbar\to M'_1 M'_2$,
respectively $\bar{A}^{M_1M_2}_d$ and $\bar{A}^{M'_1 M'_2}_d$, can be
obtained from the above amplitudes by changing the sign of the weak
phase $\gamma$.

{}From time-dependent measurements of the process $\bd(t) \to M_1
M_2$, one can obtain the following three observables:
\bea
X &\equiv & \frac{1}{2} \left( \left| A^{M_1M_2}_d \right|^2 + \left|
    \bar{A}^{M_1M_2}_d \right|^2 \right) = \calP_{ct}^2 + {\cal
  P}_{ut}^2 + 2 \calP_{ct} \calP_{ut} \cos{\Delta} \cos\gamma
~, \nn \\
Y &\equiv & \frac{1}{2} \left( \left| A_d^{M_1M_2}\right|^2 - \left|
    \bar{A}^{M_1M_2}_d \right|^2 \right) = 2 \calP_{ct} {\cal
  P}_{ut} \sin{\Delta} \sin\gamma
~, \nn \\
\ZI &\equiv & -{\rm Im}\left( e^{-2i \beta}A_d^{M_1M_2*}
  \bar{A}^{M_1M_2}_{d } \right) = \calP_{ct}^2 \sin 2\beta + 2
\calP_{ct} \calP_{ut} \cos{\Delta} \sin (2\beta + \gamma) \nn\\
& & \hskip3truein
+ \calP_{ut}^2 \sin (2\beta + 2\gamma) ~,
\label{ZPI} 
\eea
where ${\Delta}\equiv {\delta}_{ct} - {\delta}_{ut}$. One can also
define a fourth observable:
\bea
\ZR & \equiv & {\rm Re}\left( e^{-2i \beta}A_d^{M_1M_2*}
  \bar{A}^{M_1M_2}_{d } \right) = \calP_{ct}^2 \cos 2\beta + 2
\calP_{ct} \calP_{ut} \cos{\Delta} \cos (2\beta + \gamma) \nn\\
& & \hskip3truein
+ \calP_{ut}^2 \cos (2\beta + 2\gamma) ~.
\eea
Given that the width difference between $\bd$ and $\bdbar$ is very
small, it is unlikely that $\ZR$ can be measured experimentally.
However, note that $\ZR$ is not independent of the other three
observables:
\beq
\ZR^2 = X^2 - Y^2 - \ZI^2 ~.
\eeq
Thus, one can obtain $\ZR$ from measurements of $X$, $Y$ and $\ZI$, up
to a sign ambiguity. It is also useful to further define ``rotated''
observables:
\bea
\tildeZI & \equiv & \ZI \cos 2\beta - \ZR \sin 2\beta \nn\\
& = & \calP_{ut}^2 \sin 2\gamma + 2 \calP_{ct} \calP_{ut}
\cos{\Delta} \sin \gamma ~, \nn\\
\tildeZR & \equiv & \ZI \sin 2\beta + \ZR \cos 2\beta \nn\\
& = & \calP_{ut}^2 \cos 2\gamma + \calP_{ct}^2 + 2 \calP_{ct}
\calP_{ut} \cos{\Delta} \cos \gamma ~.
\eea
Assuming that $\beta$ is known independently (e.g.\ from the CP
asymmetry in $B \to J/\psi K_s$), we can obtain $\tildeZI$ and
$\tildeZR$ from measurements of $\bd(t) \to M_1 M_2$. The observables
$X'$, $Y'$, etc.\ for the second process $\bd \to M'_1 M'_2$ can be
defined similarly using the ``primed'' parameters of
Eq.~(\ref{amptilde}).

Note that, apart from the CP phase $\beta$, the three independent
observables $X$, $Y$ and $\ZI$ depend on four unknowns: ${\cal
  P}_{ut}$, $\calP_{ct}$, $\Delta$, $\gamma$. Thus, one cannot
obtain CP-phase information from the process $\bd(t) \to M_1 M_2$
alone. However, the above equations can be solved to yield ${\cal
  P}_{ut}$ and $\calP_{ct}$ as functions of $\gamma$:
\bea
\calP_{ut}^2 & = & {\tildeZR - X \over \cos 2\gamma - 1} ~, \nn\\
\calP_{ct}^2 & = & {\tildeZR \cos 2\gamma + \tildeZI \sin 2\gamma -
  X \over \cos 2\gamma - 1} ~.
\label{sprocess}
\eea
Thus, combining both processes, one can write
\beq
{r_t \over r_t'} \equiv { \calP_{ct}/\calP_{ut} \over
  \calP'_{ct}/\calP'_{ut} } = \sqrt{ {\tildeZR \cos 2\gamma + \tildeZI \sin
    2\gamma - X \over \tildeZR' \cos 2\gamma + \tildeZI' \sin 2\gamma
    - X' } \, {\tildeZR' - X' \over \tildeZR - X}} ~.
\label{gammasolve}
\eeq
We therefore see that a prediction for $r_t/r_t'$ will allow us to
obtain $\gamma$.

As shown earlier, $r_t/r_t'$ is expected to lie in the range $0.75 \le
r_t/r_t' \le 1.34$. Although this range is far more narrow than those
found for $r_u/r_u'$ and $r_c/r_c'$, it is still very large. How does
this translate into an error on the extracted value of $\gamma$? To
examine this question, we take the true values of the theoretical
parameters to be:
\bea 
& \calP_{ct} = 1.1 ~,~~ \calP_{ut} = 1.0 ~,~~ 
\calP'_{ct} = 1.5 ~,~~ \calP'_{ut} = 1.36 ~, & \nn\\
& \Delta = 30^\circ ~,~~ \Delta' = 110^\circ ~,~~ \beta = 20^\circ ~. &
\label{inputparams1}
\eea
We also consider three values for $\gamma$: $30^\circ$, $60^\circ$ and
$80^\circ$. Given these inputs, we can calculate the values of the
experimental quantities in Eq.~(\ref{gammasolve}). Then, given a value
of $r_t/r_t'$, we can solve for $\gamma$. In all cases, we compute the
range for $\gamma$ obtained if one takes $r_t/r_t' = 1.0 \pm \Delta
r$, for several values of $\Delta r$: 0.01, 0.05, 0.1, 0.2, 0.25.

The results are shown in Table \ref{gammarange1}. (We ignore the
discrete ambiguities present in the extraction of $\gamma$ from
Eq.~(\ref{gammasolve}).) For $r_t/r_t' = 1.0 \pm 0.25$, which is
almost the full allowed range of $r_t/r_t'$, the theoretical error on
the extracted value of $\gamma$ is about 20--25\%, which is quite
large. Thus, as things stand, this method cannot be used to make a
precision measurement of $\gamma$. 

\begin{table}
\hfil
\vbox{\offinterlineskip
\halign{&\vrule#&
 \strut\quad#\hfil\quad\cr
\noalign{\hrule}
height2pt&\omit&&\omit&&\omit&\cr 
& $\gamma_{in}$ && $\Delta r$ && range of $\gamma$ & \cr
height2pt&\omit&&\omit&&\omit&\cr 
\noalign{\hrule}
height2pt&\omit&&\omit&&\omit&\cr 
& $30^\circ$ && 0.01 && $29.8^\circ$ -- $30.3^\circ$ & \cr
& \omit && 0.05 && $28.8^\circ$ -- $31.4^\circ$ & \cr
& \omit && 0.1 && $27.6^\circ$ -- $32.9^\circ$ & \cr
& \omit && 0.2 && $25.4^\circ$ -- $36.3^\circ$ & \cr
& \omit && 0.25 && $24.4^\circ$ -- $38.3^\circ$ & \cr
height2pt&\omit&&\omit&&\omit&\cr 
\noalign{\hrule}
height2pt&\omit&&\omit&&\omit&\cr 
& $60^\circ$ && 0.01 && $59.7^\circ$ -- $60.6^\circ$ & \cr
& \omit && 0.05 && $57.9^\circ$ -- $62.4^\circ$ & \cr
& \omit && 0.1 && $55.7^\circ$ -- $64.9^\circ$ & \cr
& \omit && 0.2 && $51.6^\circ$ -- $70.1^\circ$ & \cr
& \omit && 0.25 && $49.6^\circ$ -- $72.9^\circ$ & \cr
height2pt&\omit&&\omit&&\omit&\cr 
\noalign{\hrule}
height2pt&\omit&&\omit&&\omit&\cr 
& $80^\circ$ && 0.01 && $79.6^\circ$ -- $80.7^\circ$ & \cr
& \omit && 0.05 && $77.6^\circ$ -- $82.7^\circ$ & \cr
& \omit && 0.1 && $75.0^\circ$ -- $85.4^\circ$ & \cr
& \omit && 0.2 && $69.9^\circ$ -- $89.3^\circ$ & \cr
& \omit && 0.25 && $67.4^\circ$ -- $93.5^\circ$ & \cr
height2pt&\omit&&\omit&&\omit&\cr 
\noalign{\hrule}}}
\caption{The range for $\gamma$ calculated from Eq.~(\protect\ref{gammasolve})
  assuming that $r_t/r_t' = 1.0 \pm \Delta r$, for the input
  parameters of Eq.~(\protect\ref{inputparams1}), and for three input
  values of $\gamma$ ($\gamma_{in} = 30^\circ$, $60^\circ$,
  $80^\circ$).}
\label{gammarange1}
\end{table}

Of course, if the theoretical uncertainty on $r_t/r_t'$ could be
improved, this would greatly reduce the error on $\gamma$. For
example, as shown in Table \ref{gammarange1}, if the uncertainty on
$r_t/r_t'$ were 5\%, the error on $\gamma$ would only be 2--3$^\circ$,
which would be quite acceptable. One possibility for reducing this
uncertainty might be to use similar final states. Recall that we have
assumed that the allowed ranges for $r_t^{(had)}$ and
$r_t^{\prime(had)}$ are completely independent. That is, for each of
$r_t^{(had)}$ and $r_t^{\prime(had)}$, we have assumed that the
momentum transfer $q^2_{av}$ and the nonfactorizable correction $x$
may take completely different values in the two decay processes.
However, as noted previously, it is quite likely that the momentum
transfer $q^2_{av}$ and the nonfactorizable correction $x$ will take
comparable values in two similar processes, which will greatly reduce
the allowed range of $r_t/r_t'$.

Unfortunately, this does not help to reduce the error on $\gamma$.
Suppose that, instead of the values given in Eq.~(\ref{inputparams1}),
the theoretical parameters take the following values:
\bea 
& \calP_{ct} = 1.1 ~,~~ \calP_{ut} = 1.0 ~,~~ 
\calP'_{ct} = 1.15 ~,~~ \calP'_{ut} = 1.05 ~, & \nn\\
& \Delta = 30^\circ ~,~~ \Delta' = 40^\circ ~,~~ \beta = 20^\circ ~. &
\label{inputparams2}
\eea
Note that the primed and unprimed parameters are similar to one
another, so these could represent $B$ decays to two similar final
states. We again consider three values for $\gamma$: $30^\circ$,
$60^\circ$ and $80^\circ$. As before, we can calculate the extracted
value of $\gamma$ using Eq.~(\ref{gammasolve}) for $r_t/r_t' = 1.0 \pm
\Delta r$, with $\Delta r=$ 0.01, 0.02, 0.03, 0.05. The results are
shown in Table \ref{gammarange2}. Regardless of the true value of
$\gamma$, if the theoretical uncertainty on $r_t/r_t'$ is greater than
about 2\%, the error on the extracted value of $\gamma$ is enormous,
particularly as regards the upper limit. In fact, even for an
uncertainty of $\Delta r = 2\%$, the corresponding error on $\gamma$
is at least 15\%, which is still large. 

\begin{table}
\hfil
\vbox{\offinterlineskip
\halign{&\vrule#&
 \strut\quad#\hfil\quad\cr
\noalign{\hrule}
height2pt&\omit&&\omit&&\omit&\cr 
& $\gamma_{in}$ && $\Delta r$ && range of $\gamma$ & \cr
height2pt&\omit&&\omit&&\omit&\cr 
\noalign{\hrule}
height2pt&\omit&&\omit&&\omit&\cr 
& $30^\circ$ && 0.01 && $27^\circ$ -- $32^\circ$ & \cr
& \omit && 0.02 && $25^\circ$ -- $39^\circ$ & \cr
& \omit && 0.03 && $24^\circ$ -- $90^\circ$ & \cr
& \omit && 0.05 && $22^\circ$ -- $107^\circ$ & \cr
height2pt&\omit&&\omit&&\omit&\cr 
\noalign{\hrule}
height2pt&\omit&&\omit&&\omit&\cr 
& $60^\circ$ && 0.01 && $54^\circ$ -- $64^\circ$ & \cr
& \omit && 0.02 && $51^\circ$ -- $74^\circ$ & \cr
& \omit && 0.03 && $49^\circ$ -- $112^\circ$ & \cr
& \omit && 0.05 && $45^\circ$ -- $121^\circ$ & \cr
height2pt&\omit&&\omit&&\omit&\cr 
\noalign{\hrule}
height2pt&\omit&&\omit&&\omit&\cr 
& $80^\circ$ && 0.01 && $73^\circ$ -- $84^\circ$ & \cr
& \omit && 0.02 && $69^\circ$ -- $95^\circ$ & \cr
& \omit && 0.03 && $66^\circ$ -- $125^\circ$ & \cr
& \omit && 0.05 && $62^\circ$ -- $130^\circ$ & \cr
height2pt&\omit&&\omit&&\omit&\cr 
\noalign{\hrule}}}
\caption{The range for $\gamma$ calculated from Eq.~(\protect\ref{gammasolve})
  assuming that $r_t/r_t' = 1.0 \pm \Delta r$, for the input
  parameters of Eq.~(\protect\ref{inputparams2}), and for three input
  values of $\gamma$ ($\gamma_{in} = 30^\circ$, $60^\circ$,
  $80^\circ$).}
\label{gammarange2}
\end{table}

An examination of Eq.~(\ref{gammasolve}) reveals why the use of
similar final states does not help to reduce the error on the
extracted value of $\gamma$. If the two final states are similar, one
expects that the experimental observables will also be similar, i.e.\ 
$X \simeq X'$, $Y \simeq Y'$, etc. However, in the limit that these
sets of observables are equal to one another, Eq.~(\ref{gammasolve})
becomes independent of $\gamma$, and reduces to the tautology $1 = 1$.
Thus, although the error on $r_t/r_t'$ may be reduced for similar
final states, the KLY method breaks down if the states are too
similar. The net effect is that the error on $\gamma$ for similar
final states is actually larger than for final states which are quite
different.

{}From this analysis, we conclude that, for similar final states, we
really need a theoretical uncertainty of $\Delta r \lsim 1\%$ in order
to be able to extract $\gamma$ with a reasonable precision.
Unfortunately, with our present knowledge of hadron physics, it does
not seem possible to definitively establish that $\Delta r \lsim 1\%$
for a particular pair of $B$ decays. We therefore conclude that the
KLY method works best for pairs of final states whose hadronic
parameters are quite different. However, in this case $\gamma$ can
only be measured with a precision of $\pm 20$--25\%, given our current
understanding of hadronic physics.

We must emphasize that our calculations have all been done within the
framework of factorization, in which there is still a great deal of
hadronic uncertainty. However, there is an enormous amount of ongoing
work on exclusive hadronic $B$ decays. It may well be that, in a
couple of years, we will understand hadronic $B$ decays well enough to
theoretically predict the value of $r_t/r_t'$ for exclusive states
with reasonable precision, even for very different final states. If
this happens, then the KLY method can be used to obtain $\gamma$.

In practice, however, the KLY method will probably be turned around,
and will be used to learn about hadronic physics. That is, given an
independent measurement of $\gamma$, along with measurements of two
$b\to d$ penguin decays, Eq.~(\ref{gammasolve}) can be used to obtain
$r_t/r'_t$. This information will allow us to test the various models
of hadronic $B$ decays.

To summarize, we have re-examined the technique proposed in
Ref.~\cite{KLY} for measuring $\beta$ (the KLY method). Their original
idea is the following: consider the two $b\to d$ penguin decays
$\bd(t) \to K^0\kbar$ and $\bs(t) \to \phi \ks$. If one assumes that
one knows the value of the ratio of two matrix elements in the first
process divided by the ratio of the corresponding matrix elements in
the second process, the CP angle $\beta$ can be obtained via
time-dependent measurements of these decays.

In this paper we have pointed out that there is an ambiguity inherent
in this approach, namely that there are three possible ratios of
matrix elements one can use. Depending on which assumption one makes,
the KLY method can be used to extract $\alpha$, $\beta$ or $\gamma$.
We have used factorization to show that, in fact, it is the assumption
which allows $\gamma$ to be obtained which is the most accurate. Thus,
it seems that the KLY method can be used to obtain this CP angle.

Unfortunately, $\gamma$ can not be extracted very precisely. There is
a theoretical uncertainty in the value of the ratios of matrix
elements for the two decays, which we estimate to be as large as
25--30\%. This leads to an error on $\gamma$ of about 20--25\%, which
is substantial. Given our current understanding of hadronic physics,
it does not appear possible at present to reduce this error.

\section*{\bf Acknowledgments}

C.S.K. thanks D.L. for the hospitality of the Universit\'e de
Montr\'eal, where some of this work was done. The work of C.S.K. was
supported in part by the KRF Grants, Project No. 2000-015-DP0077. The
work of A.D. and D.L. was financially supported by NSERC of Canada.


\end{document}